Measurement of Dispersion of a Clinical Proton Therapy Beam

Satomi Shiraishi, PhD[†], Michael G. Herman, Ph.D and Keith M. Furutani, PhD[†]*

Department of Radiation Oncology, Mayo Clinic, Rochester, Minnesota.

*Corresponding Author.  Department of Radiation Oncology, Mayo Clinic, 200 First St SW, Rochester, MN 55905 (Furutani.Keith@mayo.edu).

† These authors contributed equally to this work.

**Abstract**

**Purpose:** To measure dispersion of a clinical proton therapy beam

**Methods and Materials:** The proton center at our institution uses a Hitachi PROBEAT V synchrotron that delivers beams with energies ranging from 70 to 230 MeV to five treatment rooms, four of which have gantries and one fixed beam. For this study, the magnetic field strengths of all optics in the transport line were varied to measure the dispersion of 230 MeV beams in each of the four gantry rooms. Beam position was measured using the Spot Position Monitor (SPM), a multi-wire proportional counter (MWPC) in the nozzle approximately 0.5 meters upstream of the treatment isocenter. Measured beam positions as a function of momentum offset were fitted to a linear function to extract dispersion. In one gantry, dispersions for 70 and 140 MeV beams were also measured. Because spot position is continuously monitored during treatments and a position deviation will abort the beam delivery, the dispersion at the spot position monitor was used to set an upper limit of range uncertainty originating from momentum deviation. When dispersion was approximately zero, dose rate measurements were used to set an upper limit of momentum deviation that could reach the patient.

**Results:** Dispersion was 0.45 m, 0.48 m and 0.30 m for 70 MeV, 140 MeV, and 230 MeV beams in a gantry, respectively, corresponding to 0.3 mm, 1.0 mm and 3.5 mm maximum range uncertainty during treatments. For 230 MeV beams, dispersion ranged from 0 to 0.73 m across the four gantries, corresponding to the upper limits of range uncertainty due to momentum deviation of 1.4 to 6.2 mm (0.4 to 1.9% in fractional range uncertainty).

**Conclusions:** The measured dispersion was both energy- and gantry-dependent and ranged from 0 to 0.73 m.

# 1 Introduction

One of the advantages of proton therapy over photon therapy is spatial conformity. The finite range of the proton beam allows delivery of treatment essentially without an exit dose, resulting in reduced total energy deposited in the patient. However, the quality of proton therapy treatments relies on accurately understanding range uncertainty to properly evaluate the treated volume; underestimating range uncertainty may result in missing the tumor target due to the potential shift of the sharp distal dose fall-off, and overestimating range uncertainty may result in unnecessary irradiation of normal tissues.

Primary sources of proton range uncertainty are the certainty of material stopping power and conversion of CT Hounsfield units (HU) to stopping power. Based on the Bethe-Bloch equation, the main parameters that drive the stopping power uncertainty are physical density and mean excitation energy[1]; density variation in the material and uncertainties from mean excitation energy for various heterogeneous media further add to the stopping power uncertainty. Additionally, uncertainties associated with assigning HU to relative stopping power increases the range uncertainty; measurement of HU numbers has uncertainty due to imaging parameters such as CT calibration and spatial resolution, and the conversion from HU to stopping power has uncertainty due to photon interaction cross-sections in the material that are not proportional to proton stopping power. A summary of these range uncertainties can be found elsewhere[1]. To account for the range uncertainties during treatment planning, the standard practice at our clinic is to study the treatment plan robustness to an uncertainty in the prescribed range of 3%. Another source of range uncertainty is the proton beam momentum uncertainty. The theoretical range of a proton beam can be calculated by[2],

$$R(E_i) = \int_{E_i}^{E_f} \frac{dE}{S(E)/\rho} \tag{1}$$

where R is the range, $E_i$ and $E_f$ are the initial and final beam kinetic energy, and S(E)/ρ is the mass-stopping power as a function of energy. For 100 – 200 MeV proton beams, this equation reduces to

$$R \approx 0.00244 E_i^{1.75} \left(\frac{g}{cm^2}\right) \tag{2}$$

where $E_i$ is in MeV. Fractional change in range due to change in beam momentum $p$ is

$$\frac{dR}{R} = 1.75 \frac{dE}{E} \tag{3}$$

$$= 1.75 \frac{(\gamma + 1)}{\gamma} \frac{dp}{p}$$

where γ is the relativistic gamma. A plot of range shift as a function beam energy and momentum offset are shown in Figure 1. For 70 to 230 MeV beams, $\frac{(\gamma+1)}{\gamma} = 1.93$ to 1.80, respectively. Therefore, as a rule of thumb:

$$\frac{dR}{R} \approx 3.3 \frac{dp}{p}.$$

For the highest beam energy clinically used at our institution, 230 MeV, a 0.3% momentum deviation leads to a change in range of approximately 1% or 3 mm. It has been suggested that small momentum acceptance in a proton therapy accelerator will prohibit beams from reaching the patient when there is a considerable energy deviation[3,4]. In this technical note, we report measurements of the upper bound on momentum deviation that can reach a patient during treatment.

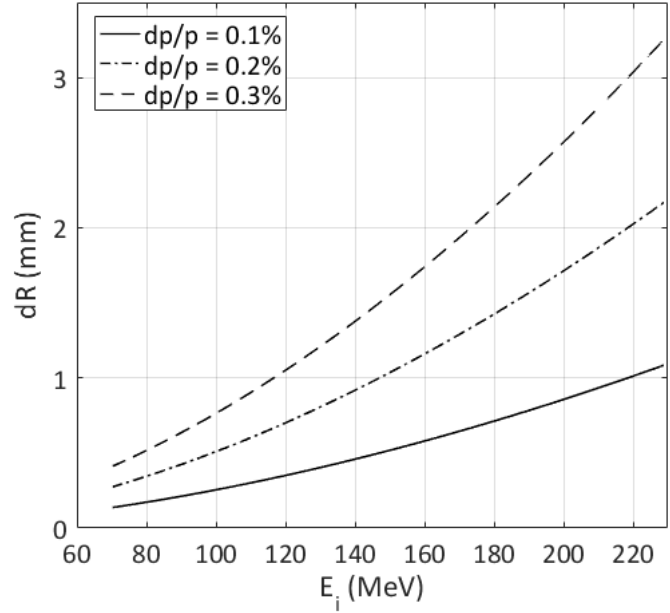

Figure 1: Change in range as a function of beam energy for three different momentum offset.

## 2    Background

Beam steering in an accelerator is performed by applying Lorentz forces, and the effects of the magnetic fields depend on particle momentum. If the particle momentum is slightly offset from a reference momentum or if the particle beams are not monochromatic, then the deflection of a particle with the wrong energy will deviate from the trajectories of a particle with a nominal momentum. The variation in the deflection in a bending magnet caused by a momentum error is called the dispersion, and consequently, the dispersion is present only in the bending plane of the dipole magnets, which is specified as the $x$-axis in this report. The magnitude of the positional deviation is characterized by[5]

$$\Delta x = D_x(s)\delta, \tag{4}$$

where $\Delta x$ is the shift from the reference trajectory, $\delta = (p - p_0)/p_0$ is the fractional momentum offset of the beam, $p_0$ is the nominal beam momentum set by the magnets, and $D_x(s)$ is the dispersion function at location s.

Typically, beam range is measured periodically as a part of beam energy quality assurance[6], but the spot position is always measured by the SPM during treatment. If dispersion is not zero at the spot position monitor, the delivered spot position with respect to a nominal position can be used to set an upper bound on momentum deviation. Range variation due to the momentum offset can then be calculated using Equation (3). Experimentally, the dispersion can be measured by changing the beam momentum $p$ or by changing the magnetic field strengths. Magnetic rigidity, $B\rho = p_0 c/qe$, defines the required magnetic field strength for a given bending radius and particle momentum, where B is the magnetic field, $\rho$ is the bending radius, $p_0$ is the reference momentum, $c$ is the speed of light, and $qe$ is the charge of particles. By changing the magnetic field strengths B, we can in effect adjust the nominal momentum $p_0$ for a given bending radius $\rho$. When a shift in beam position $\Delta x$ is measured as a function of $p$ or $p_0$, dispersion can be calculated using Equation (4) at the given location.

## 3    Materials and Methods

The proton center at our institution uses  the Hitachi PROBEAT V synchrotron that delivers beams from 70 to 230 MeV for five treatment rooms, four of which have gantries: Gantries 1 – 4. All gantries utilize pencil beam scanning delivery with an identical gantry design. Protons are extracted from the synchrotron and transported through series of quadrupole magnets and dipole magnets to the treatment isocenter. A simplified schematic of our beam transport line is shown in Figure 2. In the transport line, to first order, the source of dispersion is entirely from the series of four bending magnets indicated with blue arcs in Figure 2. Assuming that the beam is on a reference trajectory at the time of extraction from the synchrotron, the beam position deviation due to dispersion occurs only after the first bending magnet. Beam position is measured using the Spot Position Monitor (SPM) positioned at approximately 0.5 meters upstream of the treatment isocenter. The SPM is always in the beam line to measure both spot position and width during treatments. The control system will abort the beam delivery if the spot position deviates from the planned position by more than a set tolerance level at the SPM during treatments. The tolerance is comprised of both a random and systematic component[7], but we will simply use ± 1 mm for this study. Additionally, dose rate is calculated by dividing the delivered monitor unit by the delivery time for each spot, both measured by the dose monitoring ion chamber. The dose monitoring ion chamber is located immediately upstream of the SPM as indicated in Figure 2. The field delivery timer will abort a beam if the delivery takes longer than 120% of nominal delivery time during treatments. Spot position, width and beam delivery timer interlocks were all turned off during the experiment to allow measurements outside of treatment tolerances.

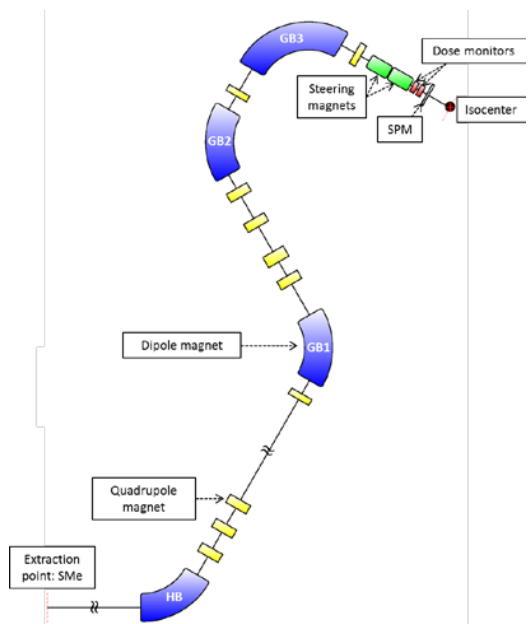

Figure 2: Simplified schematic (not to scale) of beam transport line in our proton therapy accelerator complex.

In this study, we varied the magnetic field strengths of all optics in the transport line to measure the beam dispersion. The magnet currents for these beam transport optics are linearly proportional to magnetic field strengths and were varied by $\pm0.3\%$ from the nominal values to modify the momentum offset $\delta$. For each magnet setting, 300 spots of 0.04 monitor units were delivered to the isocenter and the beam profile was measured with the SPM. For each spot, the peak of the measured profile – defined as >30% of the peak intensity – was fitted to a Gaussian distribution to extract spot position and width. Positional shift $\Delta x$ was defined as the deviation from nominal spot position which was defined as the average spot position for $\delta = 0$. The average shifts as a function of momentum offset $\delta$ were fitted to a linear function to measure the dispersion at the SPM. Dose rate was used to evaluate if the beam was clipped in the transport line. The measured dose rate was normalized by the average dose rate for the nominal momentum, $\delta = 0$. When beam clipping was suspected based on the decreased dose rate, the data points were excluded from the linear fit for the dispersion measurement. If dispersion was not zero at the SPM, changes in beam range corresponding to spot position deviation of 1 mm were calculated as an upper limit of range uncertainty due to momentum deviation during treatments. When dispersion was approximately zero at the SPM plane, dose rate measurements were used to set an upper limit of momentum deviation that could potentially reach the patient. The measurements were performed for three beam energies -- 70 MeV, 140 MeV and 230 MeV – in Gantry 4. For the highest beam energy, the measurements were obtained for all four gantries.

## 4   Results

Beam position shift $\Delta x$ at the SPM in the bending plane ($x$-axis) as a function of $\delta$ measured for three different beam energies is shown in Figure 3 (a). The error bars indicate the standard deviation of the 300 spots for the reference momentum setting and that of momentum offset summed in quadrature. Dispersions measured by linear fit to the data are summarized in Table 1. Changes in range corresponding to 1 mm positional shifts at the SPM are also listed in Table 1. For the nominal momentum, $\delta = 0$, spot position deviation for each of the 300 spots is shown in Figure 3 (b), illustrating the drift of the spot position.  The sudden change in spot position near the beginning of the spill in the 230 MeV data which is not observed in the 70 or 140 MeV data is suspected to be due to the synchrotron magnet control system[8].   The gradual drift in spot position over the course of the spill is likely due to the drift in the nominal momentum of the extracted phase space.   Measured dose rate for $\delta = -0.3\%, 0\%, 0.3\%$ are shown in Figure 4. For 70 MeV and 140 MeV, beam was clipped for $\delta = 0.3\%$, and the decreased dose rate is seen in Figure 4 (a, b). For 230 MeV, no obvious change in dose rate was observed for all momentum settings.

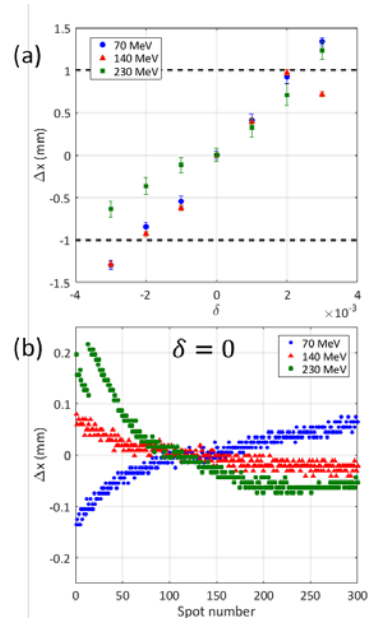

Figure 3: Spot position measurement for three different beam energies in gantry 4. (a) Average spot position for the 300 spots as a function of momentum offset, **δ**. (b) Spot position for each of the 300 spots for **δ = 0**.

Table 1: Dispersion and maximum range variation corresponding to the 1 mm positional tolerance at the SPM measured in Gantry 4.

|         | $D_x$ (m) | $dR_{max}$ (mm) | $dR_{max}/R$ |
|---------|-----------|-----------------|--------------|
| 70 MeV  | 0.45      | 0.3             | 0.8%         |
| 140 MeV | 0.48      | 1.0             | 0.7%         |
| 230 MeV | 0.30      | 3.5             | 1.1%         |

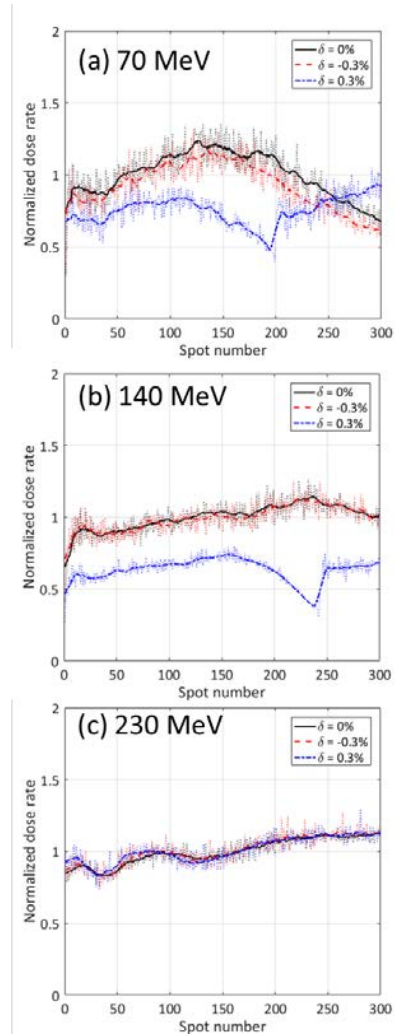

Figure 4: Normalized dose rate for each of the 300 spots for $\delta = -0.3\%, 0\%, 0.3\%$. The nominal dose rate is defined as the average dose rate for $\delta = 0\%$.

Beam position shift at the SPM as a function of $\delta$ for 230 MeV beams in all four gantry rooms are shown in Figure 5. For Gantry 2 through 4, the dispersions at the SPMs measured by the linear fit are summarized in Table 2. For Gantry 1, dispersion was approximately zero at the SPM. However, as shown in Figure 6, measured dose rate starts to decrease for $\delta \pm 0.5\%$ and decreases significantly for $\delta = \pm0.6\%$, indicating that the beam was being clipped in the transport line. For $\delta \pm 0.7\%$, the beam failed to reach the treatment room, even with the beam delivery interlocks turned off. This occurred due to non-zero dispersion in the beam line upstream of the SPM. Due to the nature of dipole optics, the location of maximum dispersion would occur between GB1 and GB2 dipole magnets for all four gantry rooms[5]. The beam is lost at this point of maximum dispersion because of the finite diameter of the beam pipe. The momentum shift of $\delta = 0.6\%$ for the 230 MeV beam corresponds to a 6.2 mm change in range or 1.9% in fractional change in range. For gantries 2 through 4, the normalized dose rate was similar to what is shown in Figure 4 (c); no obvious beam clipping was observed for $\delta \pm 0.3\%$.

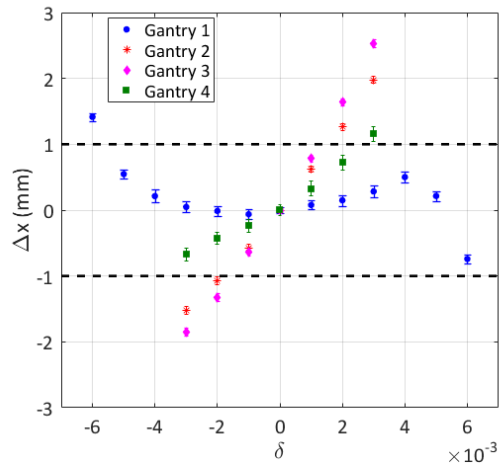

Figure 5: Spot position shifts at the SPM for 230 MeV beam in all four gantries.

Table 2: Dispersion and maximum range variation corresponding to the 1 mm positional tolerance at the SPM for 230 MeV. *Upper limit calculated for $\delta = 0.6\%$, where we observed a decrease in dose rate.

|  | $D_x$ (m) | $dR_{max}$ (mm) | $dR_{max}/R$ |
|---|---|---|---|
| Gantry 1 | -- | 6.2* | 1.9% |
| Gantry 2 | 0.58 | 1.8 | 0.5% |
| Gantry 3 | 0.73 | 1.4 | 0.4% |
| Gantry 4 | 0.30 | 3.5 | 1.0% |

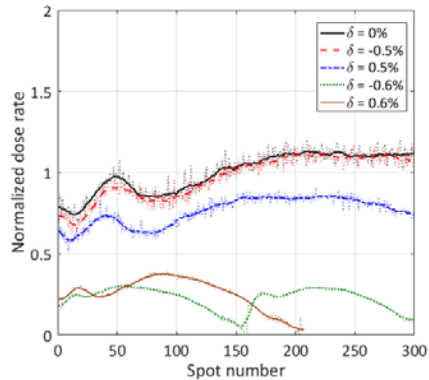

Figure 6: Measured dose rate in Gantry 1 for $\delta = 0\%, \pm 0.5\%, \pm 0.6\%$.

## 5   Discussion

In this study, we measured dispersion at the SPM for our highest beam energy in all four gantries, and we measured the dispersions of three beam energies in Gantry 4. Dispersion was both gantry- and beam energy-dependent. Spot position tolerance of $\pm 1$ mm sets an upper limit of range uncertainty originating from momentum deviation: 0.3 mm for the 70 MeV and 3.5 mm for the 230 MeV beams in Gantry 4. Furthermore, dispersion at the SPMs varied across the four gantries. For Gantries 2 through 4, upper limits of range uncertainty due to momentum deviation were 1.4 to 3.5 mm or 0.4 to 1.0% in fractional range uncertainty. For Gantry 1, where dispersion was approximately zero at the SPM, dose rate measurements indicated that momentum offset of $\delta = 0.6\%$ clips the beam, which would abort the delivery based on the field delivery timer. The momentum shift of 0.6% for the 230 MeV beam corresponds to a 6.2 mm change in range. The upper limit of range uncertainty is likely an overestimate because the analysis assumes that spot position deviation is solely caused by momentum deviation. However, understanding dispersion at the SPM and the momentum acceptance of the transport line is a valuable consideration for the patient safety.

Ideally, spot position deviation would be measured at a location with a large dispersion function to achieve high sensitivity to deviations in momentum. Since gantries are generally designed to have zero or very small dispersion at the treatment isocenter, measuring spot position closer to the treatment isocenter would likely decrease the sensitivity of this dispersion test. Continuous monitoring of beam position at the largest dispersion location in the accelerator complex using a passive beam position monitor would be ideal. However, in a capacitive coupling device, low beam currents in our transport line -- on the order of a few hundred nano-amps -- would generate only a small signal that would likely be lost in the noise[9]. Another possibility to decrease the variation in particle range potentially reaching the patient is to reduce the diameter of the beam pipe at the position of largest dispersion. This option seems attractive because then one could eliminate the SPM to reduce multiple scattering in the nozzle and therefore decrease the spot size to the patient. However, one would likely still wish to retain the SPM to measure the spot size and to have a confirmation of the spot scanning magnetic fields.

Finally, it is worth noting that there are accelerators with a large momentum acceptance, such as fixed-field alternating gradient accelerators, which would be less sensitive to momentum deviation by design[10,11]. For these accelerators, assuring the beam energy reaching the patient would still be necessary to ensure patient safety.

## 6   Conclusions

We measured beam dispersion in our clinical proton therapy system 0.5 m upstream of treatment isocenter. The dispersion was both beam energy- and gantry-dependent. Measured dispersion ranged from 0 – 0.73 m across the four gantries for 230 MeV beams.

**Acknowledgement**

Authors would like to thank Takuya Nomura of the Particle Therapy Systems Design Department of Hitachi Works Inc. for his assistance in these measurements and numerous helpful discussions.